\documentclass[14pt, prc, onecolumn, nofootinbib,preprint,superscriptaddress]{revtex4-2}

\usepackage[T1]{fontenc}
\usepackage{amsmath,amssymb}
\usepackage{epsfig}
\usepackage{float}
\usepackage{subfig}

\usepackage{setspace}

\usepackage{graphicx}
\usepackage[colorlinks,citecolor=blue]{hyperref}
\usepackage{color}

\begin{document}
	\title{\textcolor{black}{Comparative Analysis of Single Charged Particle Production in Proton-Carbon Interactions at High Energies} }
		
	\author{R Lalnuntluanga}
	\email{tluangaralte.phy@gmail.com}
	\affiliation{Department of Physics, Indian Institute of Technology Hyderabad, Kandi, Sangareddy 502285, Telangana, India}
	
    \author{A Giri}
    \email{giria@phy.iith.ac.in}
    \affiliation{Department of Physics, Indian Institute of Technology Hyderabad, Kandi, Sangareddy 502285, Telangana, India}

\setstretch{1.5}

\begin{abstract}
 The generation of charged pions, kaons, and protons from proton beam incident on Carbon at 31 GeV/c is calculated and compared with the results of the NA61/SHINE experiment. Predictions of the single charged particle yield by proton off the Carbon target is calculated using the Giessen Boltzmann-Uehling-Uhlenbeck (GiBUU) model and compared with the recent data of the EMPHATIC hadron scattering and production experiment for incident proton beam energies of 20, 30, and 120 GeV/c within $\pm$20 mrad with respect to the beam particle. We present the analysis for the single and double differential cross-sections for the produced particles at various scattering angles and conduct a comprehensive comparison with the experimental data. These studies show significant agreement with the measured data.   
\end{abstract}

\maketitle
\noindent

%


\newpage
\section{Introduction}

\label{sec:1}
Accurate measurements of hadron interactions are critical to minimize the modeling uncertainty associated with neutrino production in both accelerator-based and atmospheric neutrino experiments. These measurements should span two orders of magnitude, ranging from 1 to 100 GeV/c of incident particle momenta. The dominant source of uncertainty in numerous neutrino measurements, such as those pertaining to neutrino-nucleus cross-sections, sterile neutrino searches, and CP violation in atmospheric neutrinos, is the uncertainty in the neutrino flux. The upcoming Hyper-Kamiokande \cite{Hyper-Kamiokande:2018ofw} and DUNE \cite{DUNE:2020lwj} projects mark the beginning of a precision era for long-baseline neutrino experiments. In these experiments, the energy dependence of the neutrino flux and cross-section are among the most challenging systematic uncertainties from the knowledge of the ongoing experiments \cite{PhysRevD.96.092006,PhysRevD.98.032012}. The hadrons produced in proton interactions within the nuclei decay, and produce neutrinos. Measuring the neutrino beam flux in terms of energy is a challenging and time-consuming task. Therefore, a-priori predictions of the neutrino flux are made using Monte Carlo simulations based on hadron interactions and decays\cite{PhysRevD.94.092005,Bodek:2012uu}. However, this approach is limited due to insufficient hadron cross-section data and significant errors, and hadron interactions are the main source of systematic uncertainty in the prediction of neutrino flux \cite{PhysRevD.94.092005,Bodek:2012uu}.  To reduce the uncertainties, phenomenological models are used to interpolate and extrapolate the hadron interaction cross-section, but this introduces additional uncertainties. Measurements of hadron interactions data are employed to constrain or scale the models to enhance the accuracy of the neutrino flux prediction. 

This paper presents a comparison between the calculations made using the GiBUU transport model and the data obtained from the NA61/SHINE experiment\cite{NA61SHINE:2015bad}. Previous studies \cite{Gallmeister:2009ht} show a good agreement of the GiBUU model with the HARP experiment\cite{HARP:2005clh,HARP:2008jan,HARP:2008sqs}, and also shows the predictions for the NA61/SHINE experiment\cite{Laszlo:2009vg,NA61:2008uqu,NA61:2014lfx,NA61SHINE:2011dsu}.  These studies \cite{Gallmeister:2009ht} compared the experimental data on inclusive pion production using both pion and proton beams directed at various nuclear targets with beam momenta ranging from 3 to 12 GeV/c.  The study highlights the GiBUU model's unique capability to accurately model the entire energy range covered by the HARP experiment. The results indicate that the GiBUU model provides the most accurate description of the experimental data when pion beams are used. The model also shows a very good level of agreement across the entire energy range using proton beams, although it struggles to accurately predict outcomes at the very forward and very backward angles. These discrepancies are attributed to the effect of the Final State Interactions (FSI), as comparisons with data from elementary collisions. The HADES experiment analysis \cite{HADES:2014lrq} also report the elementary p+p data and proton-niobium collisions reproduced using modified GiBUU.

This article presents the GiBUU calculations for proton beams at 31 GeV/c momentum and compared with the NA61/SHINE experimental data \cite{NA61SHINE:2015bad}. We then extend our analysis with proton beam momenta at 20, 30, and 120 GeV/c and the results are compared with the data of the EMPHATIC experiment \cite{EMPHATIC:2019xmc,EMPHATIC:2021num}.

This paper is organized as follows:  The experimental details of the NA61/SHINE and EMPHATIC experiment are described in Section \ref{sec:2} and \ref{sec:3}, respectively. In Section \ref{sec:4}, we describe the GiBUU theoretical model with a detailed description of the event generator. The results obtained from our analysis and comparison with the data are presented in Section \ref{sec:5}  and conclusions are discussed in Section \ref{sec:6}.

\begin{figure}
	\centering
	\subfloat{\includegraphics[width=1.05\linewidth]{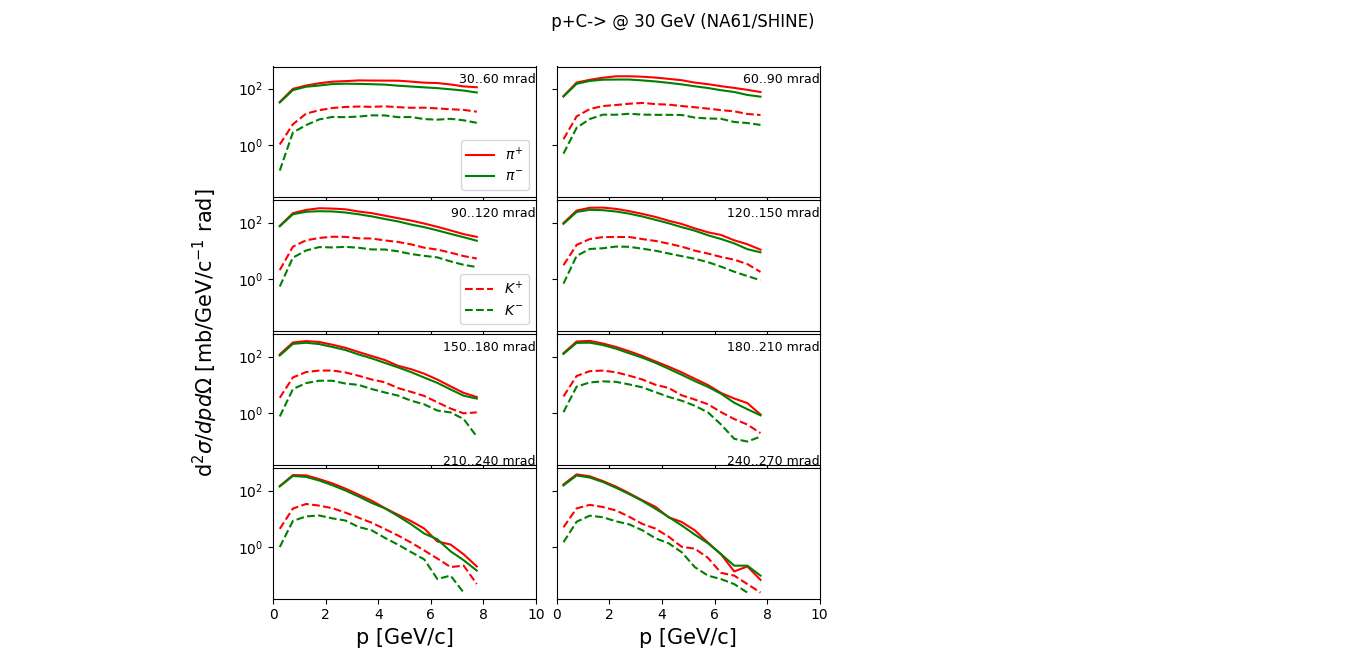} }
	
	\caption{The plot represents the cross-section $\frac{d^{2}\sigma}{dp d{\Omega}}$ for the production of $\pi^{\pm}$ (solid lines) and $K^{\pm}$ (dashed lines) for NA61/SHINE experiment using  p+C at 30GeV/c. }
	\label{fig:1}
\end{figure}

\section{NA61/SHINE experiment}
\label{sec:2}
The NA61/SHINE experiment, located at CERN, is dedicated to an extensive physics research agenda covering diverse areas of the field \cite{Laszlo:2009vg,NA61:2008uqu,NA61:2014lfx,NA61SHINE:2011dsu}. The NA61/SHINE setup was inherited from its precursor, the NA49 exeriment\cite{NA49:1999myq}. The detector comprises a system of five Time Projection Chambers (TPCs). Two of these TPCs, namely VTPC-1 and VTPC-2, are positioned within the magnetic field generated by two superconducting dipole magnets. Downstream and symmetrically aligned with respect to the beamline, there are two additional TPCs known as MTPC-L and MTPC-R. To cover the very-forward region, an extra small TPC called the GAP TPC (GTPC) is placed between VTPC-1 and VTPC-2.

The Time Projection Chambers (TPCs) are filled with a mixtures of gas consisting of Ar (Argon) and CO2 (Carbon dioxide) in specific ratios. For the Vertex TPCs (VTPCs) and the GAP TPC (GTPC), the gas mixture comprises of 90\% Ar and 10\% CO2. On the other hand, the Main TPCs (MTPCs) are filled with a gas mixture containing 95\% Ar and 5\% CO$_{2}$. A dipole magnet of the magnetic field of 1.14 Tm is used. More details about the experimental setup and data taking can be found ref\cite{NA61:2014lfx,NA61SHINE:2015bad}.  

\begin{figure}
	\centering
	\subfloat{\includegraphics[width=1.05\linewidth]{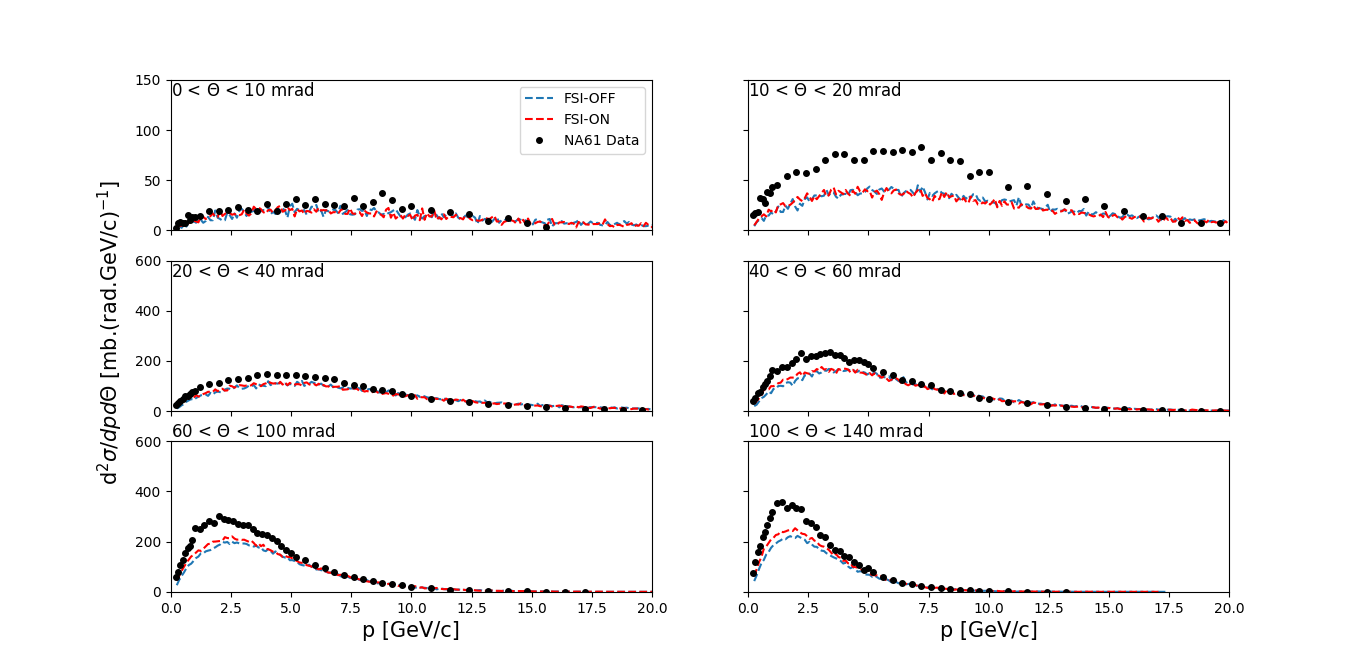} }
	
	\subfloat{\includegraphics[width=1.05\linewidth]{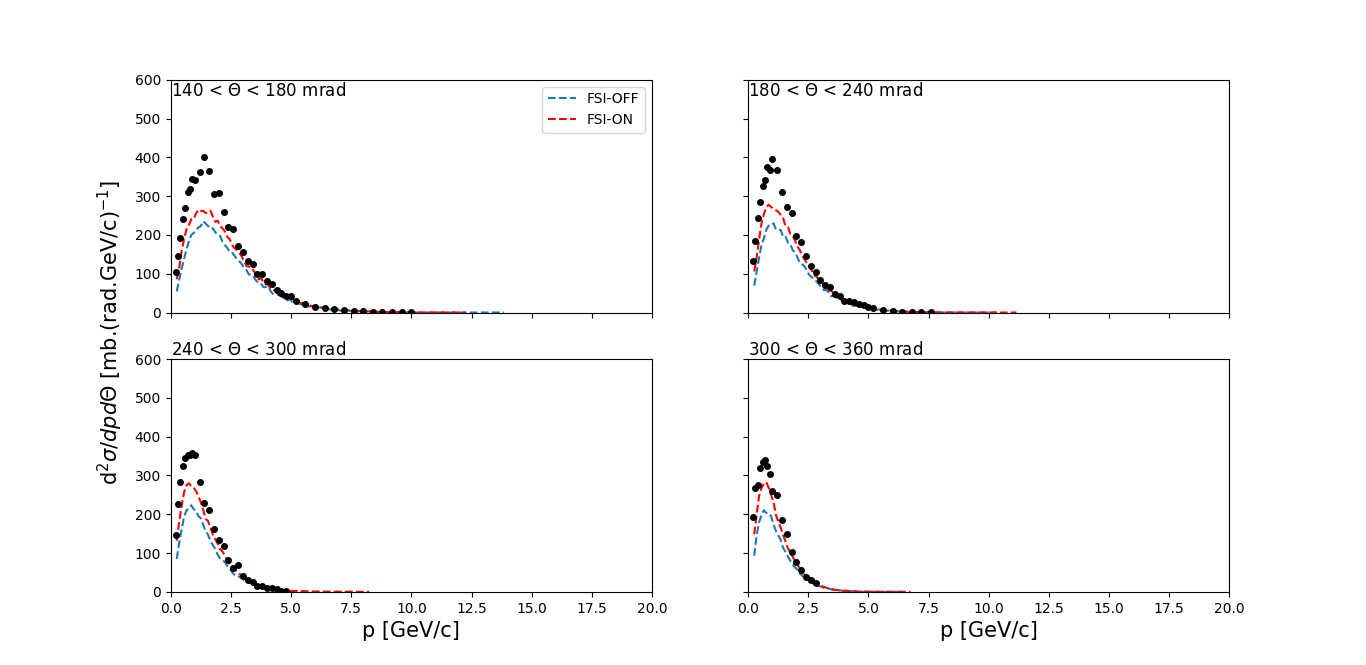} }
	\caption{The plot represents the double differential cross-section for the production of $\pi^{+}$ using GiBUU model with FSI-ON(red dashed line) and FSI-OFF(blue dashed line) compared with NA61/SHINE data(black). }
	\label{fig:2}
\end{figure}

\begin{figure}
	\centering
	\subfloat{\includegraphics[width=1.05\linewidth]{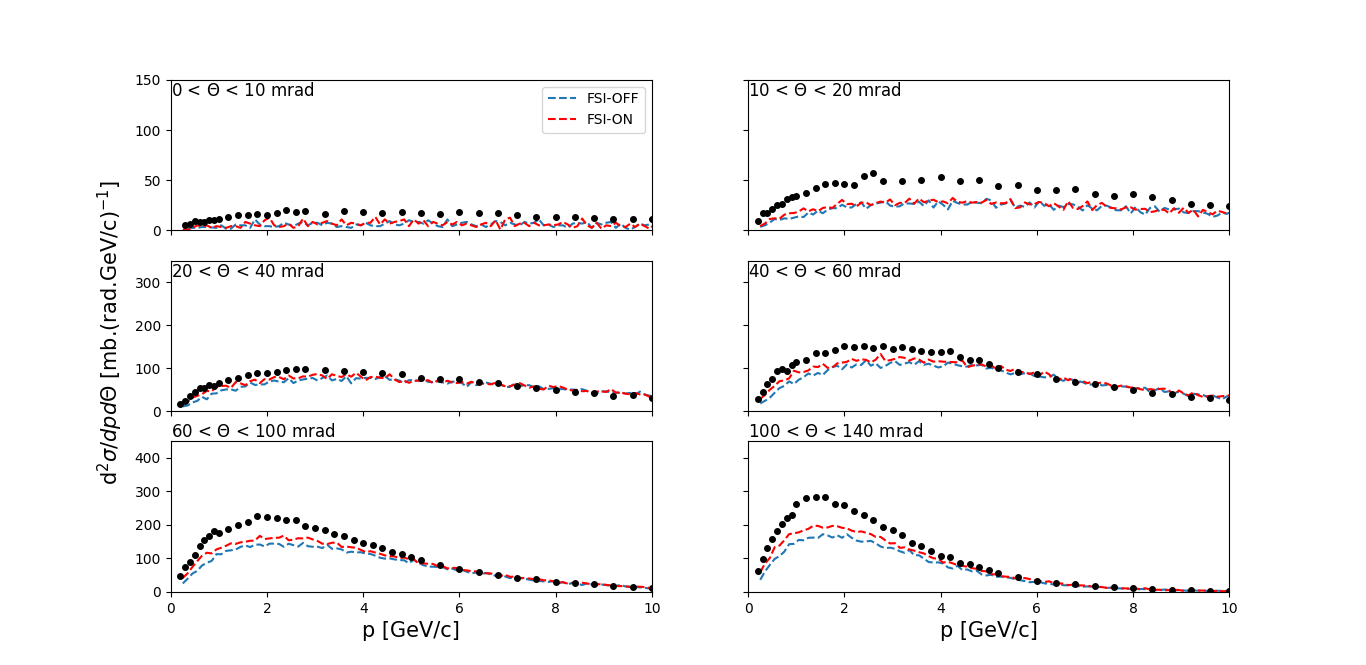} }
	
	\subfloat{\includegraphics[width=1.05\linewidth]{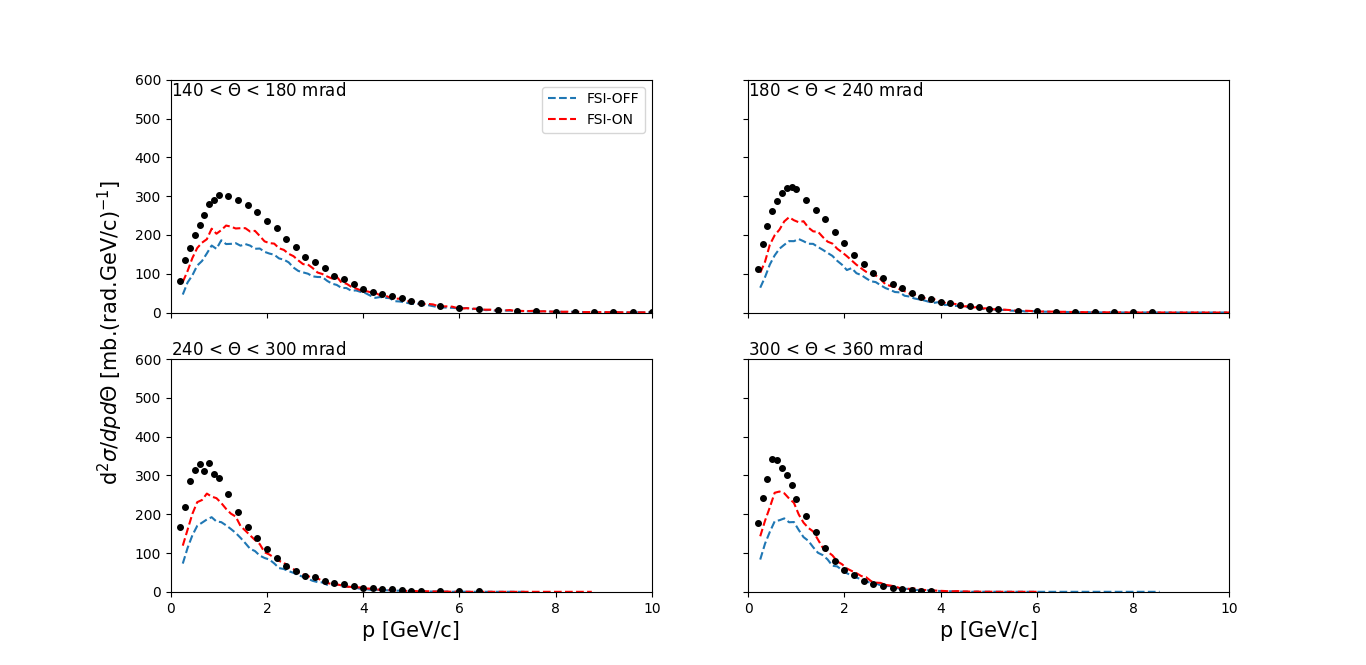} }
	\caption{The plot represents the double differential cross-section for the production of $\pi^{-}$ using GiBUU model with FSI-ON(red dashed line) and FSI-OFF(blue dashed line) compared with NA61/SHINE data(black).}
	\label{fig:3}
\end{figure}

\begin{figure}
	\centering
	\subfloat{\includegraphics[width=1.05\linewidth]{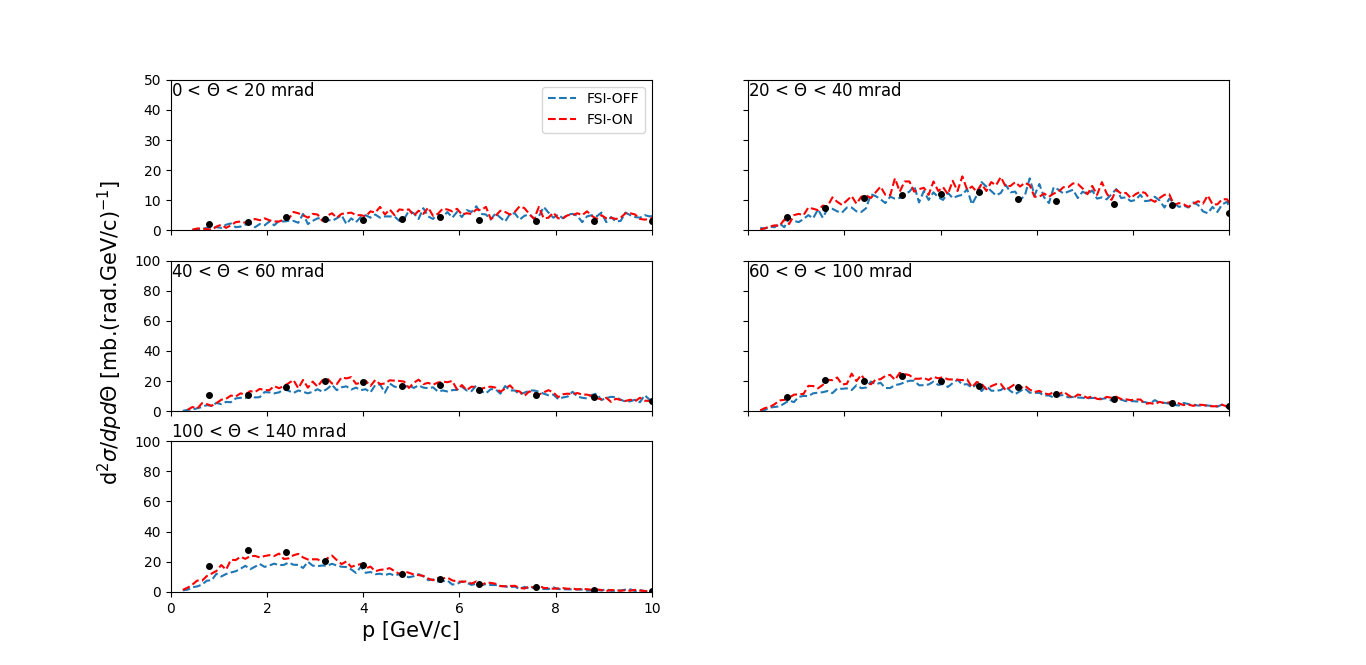} }
	
	\subfloat{\includegraphics[width=1.05\linewidth]{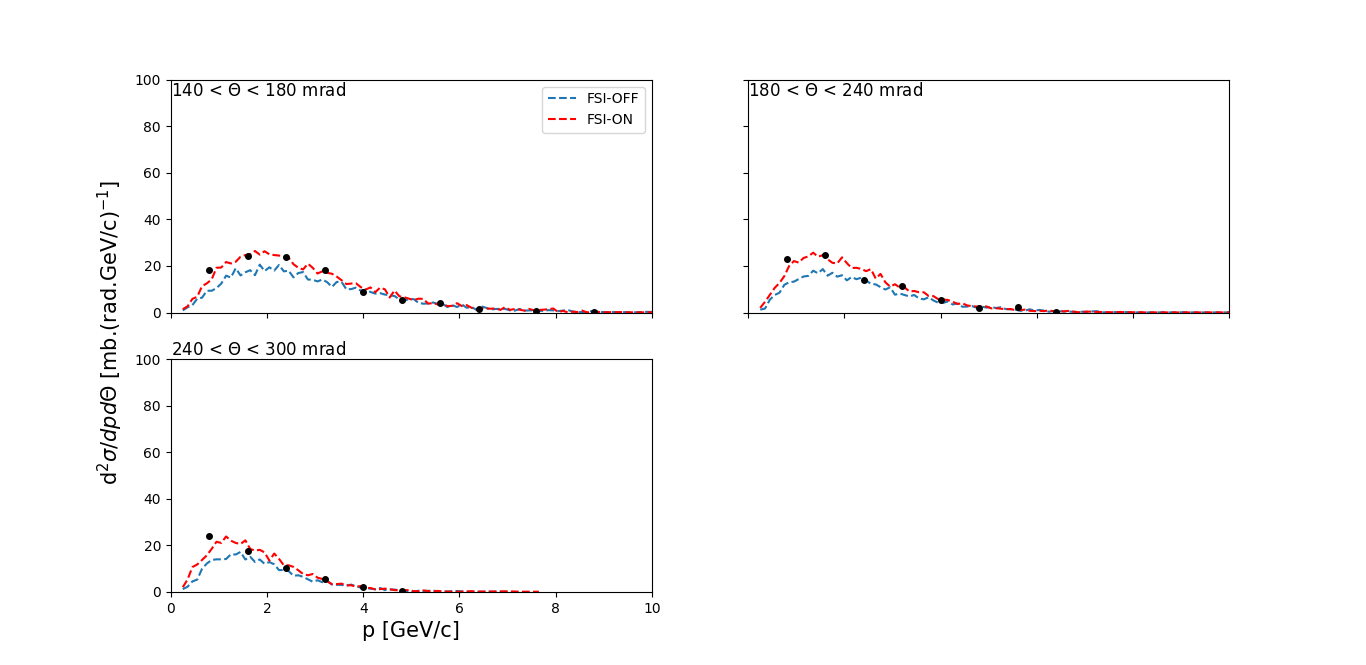} }
	\caption{The plot represents the double differential cross-section for the production of $K^{+}$ using GiBUU model with FSI-ON(red dashed line) and FSI-OFF(blue dashed line) compared with NA61/SHINE data(black).}
	\label{fig:4}
\end{figure}

\begin{figure}
	\centering
	\subfloat{\includegraphics[width=1.05\linewidth]{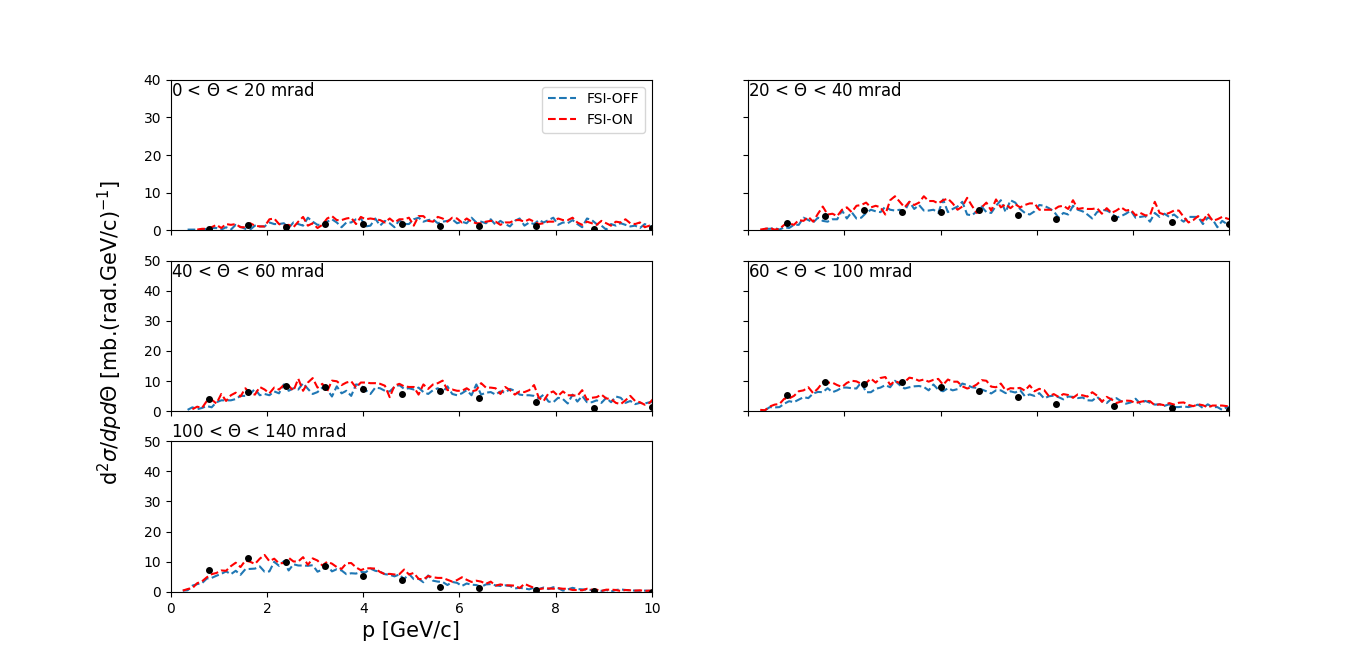} }	
	
	\subfloat{\includegraphics[width=1.05\linewidth]{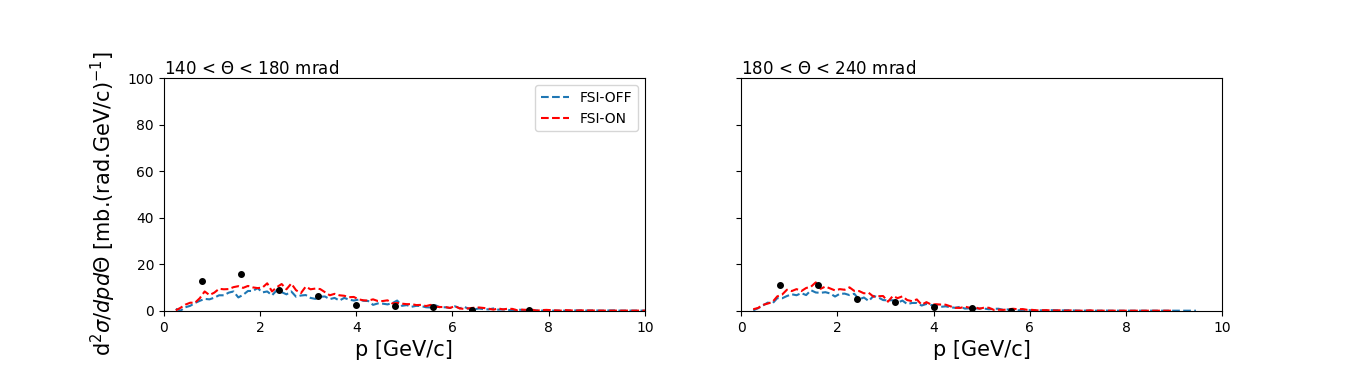} }
	\caption{The plot represents the double differential cross-section for the production of $K^{-}$ using GiBUU model with FSI-ON(red dashed line) and FSI-OFF(blue dashed line) compared with NA61/SHINE data(black).}
	
	\label{fig:5}
\end{figure}

\begin{figure}
	\centering
	\subfloat{\includegraphics[width=1.05\linewidth]{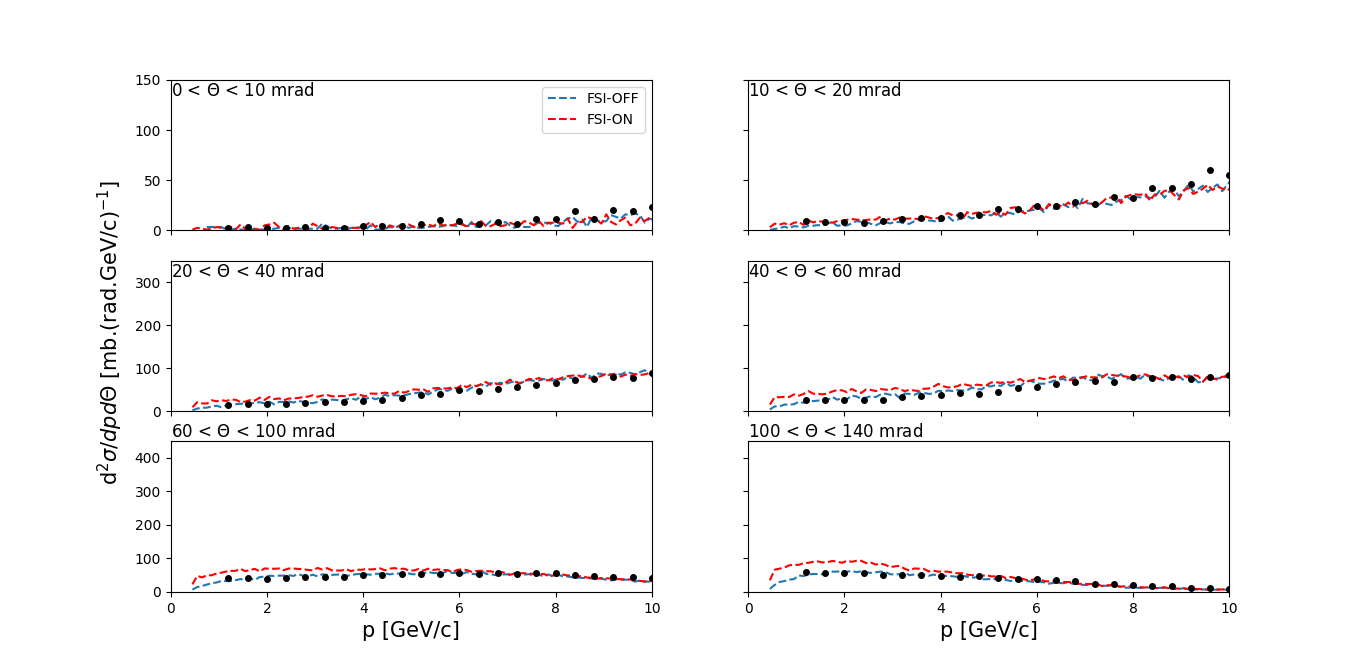} }	
	
	\subfloat{\includegraphics[width=1.05\linewidth]{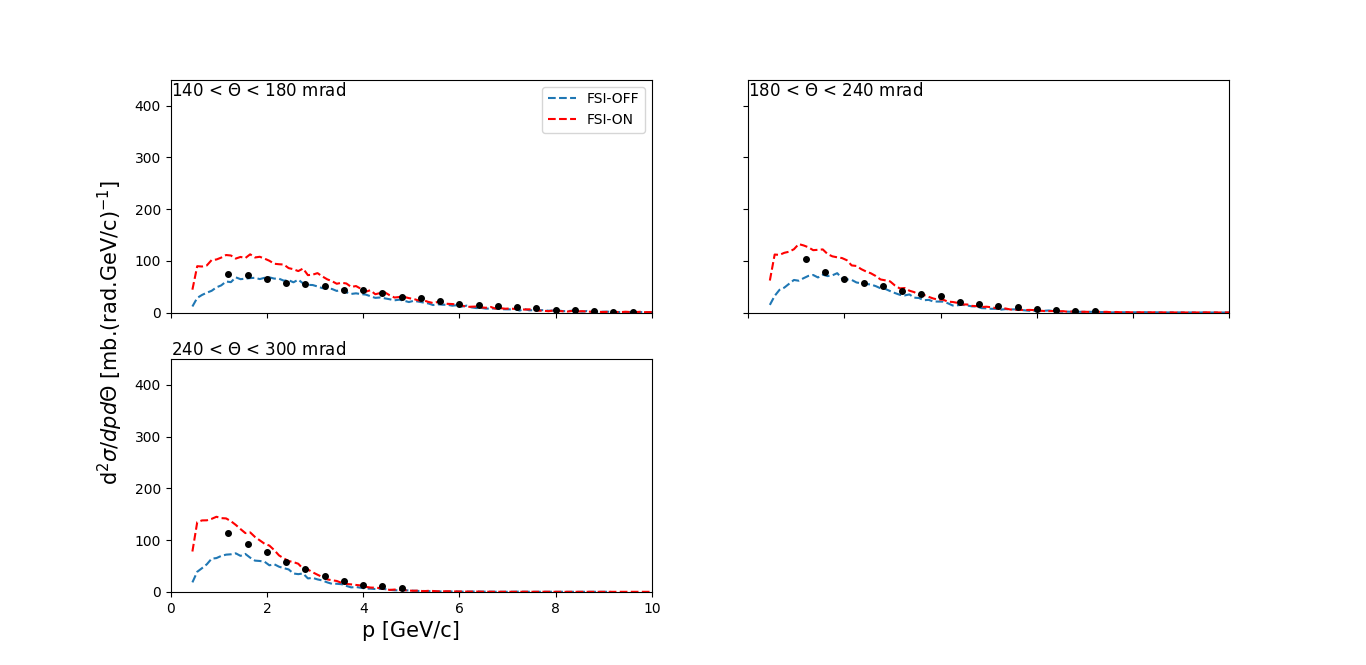} }
	\caption{The plot represents the double differential cross-section for the production of $proton$ using GiBUU model with FSI-ON(red dashed line) and FSI-OFF(blue dashed line) compared with NA61/SHINE data(black).}
	\label{fig:6}
\end{figure}

\section{EMPHATIC experiment}
\label{sec:3}
The EMPHATIC (Experiment to Measure the Production of Hadrons At a Testbeam in Chicagoland) experiment is designed to investigate hadronic interactions in the 2 - 120 GeV/c range at the FTBF (Fermilab Test Beam Facility). The compact hadron spectrometer of EMPHATIC design utilizes the silicon strip detectors' spatial resolution of $\sim$10 $\mu$m. A custom-built Halbach array of NdFeB magnets with a $\int$Bdl of about 0.25Tm serves as the dipole magnet. Gas and aerogel threshold Cherenkov detectors are used for beam particle identification, while secondary particles are identified through time-of-flight measurements in resistive plate chambers (RPCs) and measuring the Cherenkov angle in aerogel ring imaging detector (ARICH) which is based on the Belle-II design\cite{Burmistrov:2020dvn}. The separation of electrons, muons, and hadrons is enabled by a lead calorimeter at the downstream end of the experiment. The spectrometer is roughly 2m in length. More information about the experimental setup and data analysis can be found in Ref.\cite{EMPHATIC:2019xmc,EMPHATIC:2021num}.

In this paper, we consider the EMPHATIC test-beam measurement data\cite{EMPHATIC:2021num}, which include the differential cross-section for p+C$\rightarrow$X$^{\pm}$ at 20, 30, and 120 GeV/c beam momenta. A single charged particle X$^{\pm}$ is measured within $\pm$20mrad of the detector's tracking acceptances. This allows the measurements of coherent elastic and quasi-elastic interactions (forward scattering).

\section{Theoretical Model : GiBUU} 
\label{sec:4}
In this paper, we use the GiBUU model to analyze data on elementary collisions induced by baryons, mesons, (real and virtual) photons, and neutrinos on various nuclei, all within a unified framework. The GiBUU model is a transport model developed to handle energies ranging from a few MeV to $\mathcal{O}(100)$ GeV, and is structured on a modular FORTRAN 2003 code. The GiBUU-2023, patch 3 (February 21, 2024) code is used in this work and its source code can be found here\cite{Gibuu_code}. The model solves the Boltzmann-Uehling-Uhlenbeck (BUU) equations for one-particle phase-space densities, which evolved over time dynamically in the potentials and mean fields of the hadrons along with integral of collision. The collision integral is mostly governed by two-body collisions, both elastic and inelastic, which are described using a resonance information at low energies and with the Pythia\cite{Sjostrand:2014zea} at high energy. 
 For more information, such as total cross sections  plots, please refer to Ref. \cite{Gibuu_code}.

The inevitable nuclear effects, such as  Pauli blocking, nuclear shadowing and Fermi motion, are appropriately considered for in the elementary interaction between a nucleon and its environment. Next, all (pre-)hadrons produced during this interaction are transmitted in the nuclear environment using the semi-classical Boltzmann-Uehling-Uhlenbeck(BUU)  transport equation. For explaining the phenomena of formation time and color transparency, the notion of pre-hadrons are introduced where the hadrons at the hadronization process, interact with attenuate cross-section \cite{Gallmeister:2005ad,Gallmeister:2007an}. This approach ensures that every interaction, both primary and secondary, are handled consistently, resulting in a fully coupled channel transport. For beam momenta above 3 GeV/c, the Pythia method is used for elementary $p-$N or $\pi^{\pm}-$N interactions, since the collision energies are beyond the resonance zones ($\sqrt{s}_{\pi N,pN}$ = 2.8 - 4.9 GeV) and the final state particles momentum are relatively substantial, minimizing the significance of potentials of the baryons that distinguish the GiBUU description from other MC cascade models. The significance and application of the GiBUU model in neutrino physics and the measurement of hadron production are detailed here \cite{Mosel:2016cwa,Mosel:2013fxa,Larionov:2014iua,Gaitanos:2014bla,Mosel:2015oda,Mosel:2015yaa,Mosel:2016cwa,HADES:2014lrq,Mosel:2014lja}.

\section{Results}
\label{sec:5}
In order to verify our GiBUU simulation and analysis code, we have reproduced the previous results (Fig.7) of Ref.\cite{Gallmeister:2009ht} for the NA61/SHINE experiment shown in Fig.\ref{fig:1}. In this work, the default physics parameters are used without any tune\cite{Gibuu_code}. We begin our analysis with the data for the NA61/SHINE results for the proton beam at 31 GeV/c. The double differential cross-section ($\frac{d^{2} \sigma}{dp d\theta}$) depicting the production of $\pi^{+}$, $\pi^{-}$, $K^{+}$, $K^{-}$ and $p$ for all the scattering angle, are illustrated in Fig. \ref{fig:2}, \ref{fig:3}, \ref{fig:4}, \ref{fig:5}, and \ref{fig:6} respectively. From Fig. \ref{fig:2} and \ref{fig:3}, we observed that the pions ( $\pi^{+}$, $\pi^{-}$) produced by 31 GeV/c proton beam interactions with carbon in the GiBUU framework agree significantly with the experimental data in all the scattering angles. It can be seen that our calculations with the final state effect suit the data better than when the final state effect is not considered. For the forward angles (between 10 mrad and 100 mrad ), the GiBUU model prediction side with the data in the lower momentum range and slightly deviates with the increase in momentum and assent in the higher momentum tail. Further with the more forward angles ($\leqslant$ 10 mrad), our calculations accord with the data well. It can also be noted that the effect of the final state interaction slightly favours the data in the scattering angles ($\geqslant$ 20 mrad) while our calculations matched the data significantly in the backward angles (i.e $\geqslant$ 100 mrad).

From the differential cross-section for the production of the charged Kaon ($K^{+}$, $K^{-}$) shown in Fig. \ref{fig:4} and \ref{fig:5}, we can notice that our calculations agreed very well with the NA61/SHINE measurement when the final state effect is considered, in both the forward and backward angles. In case of the proton production, the GiBUU model predicted substantially well for all the angles as shown in Fig. \ref{fig:6}. In contrast to the pions and kaons production, the produced proton satisfies the measurement when the final state interaction is off.

\begin{figure}
	\centering
	\subfloat{\includegraphics[width=0.50\linewidth]{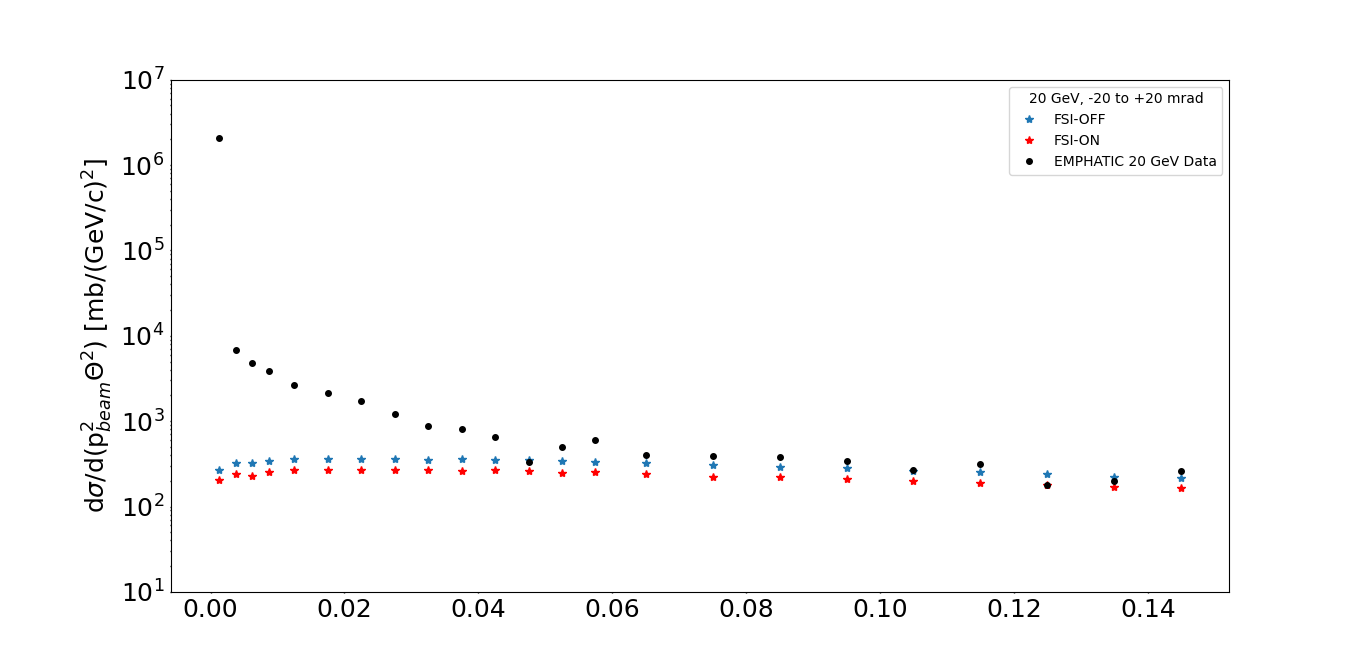} }	
	\subfloat{\includegraphics[width=0.50\linewidth]{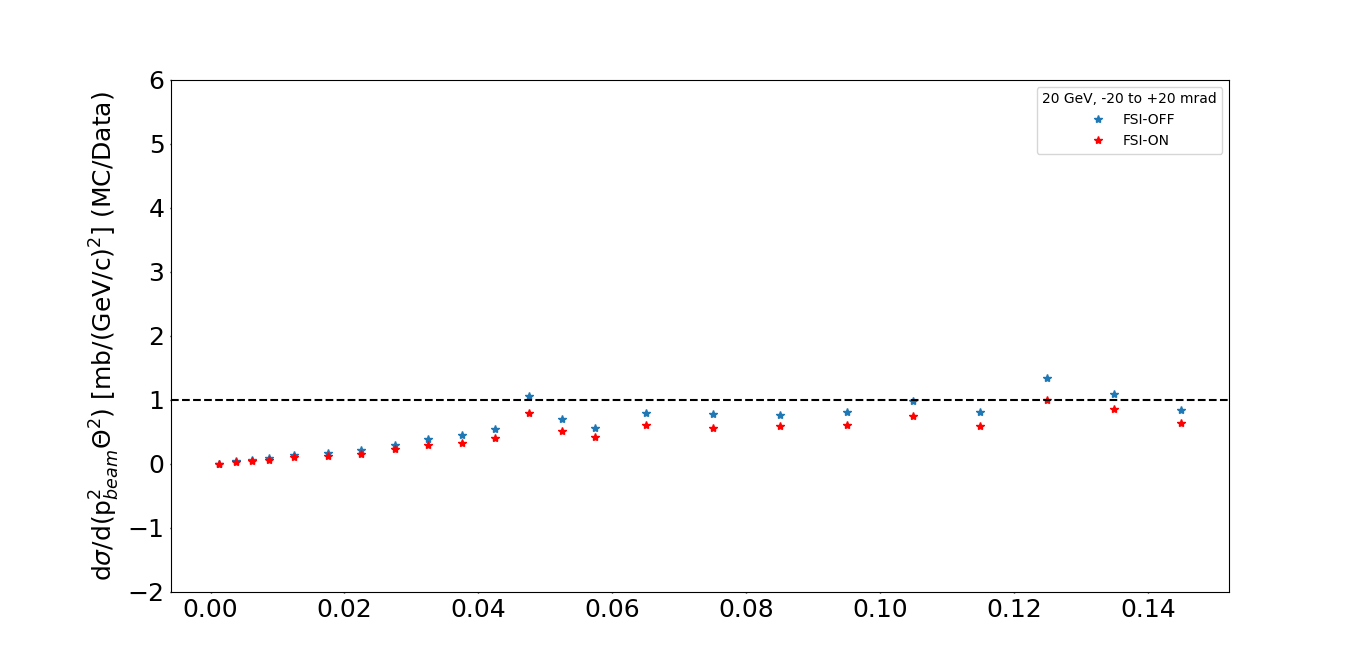} }
	
	\subfloat{\includegraphics[width=0.50\linewidth]{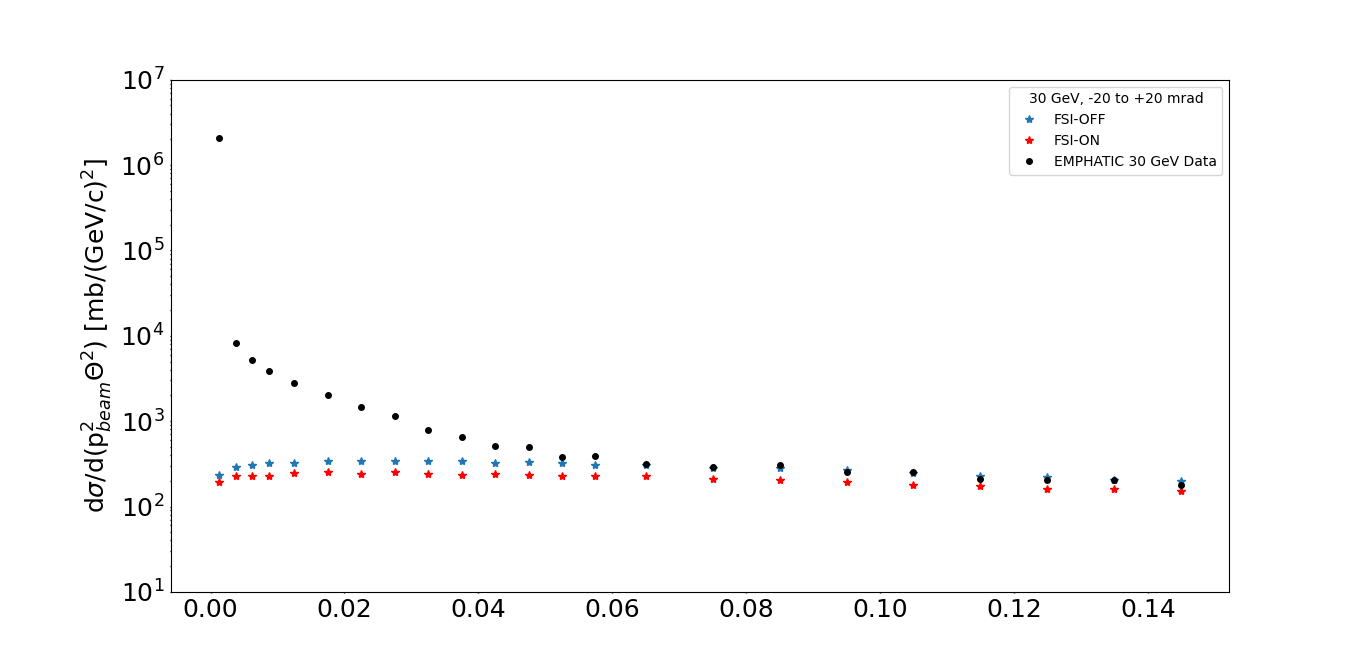} }	
	\subfloat{\includegraphics[width=0.50\linewidth]{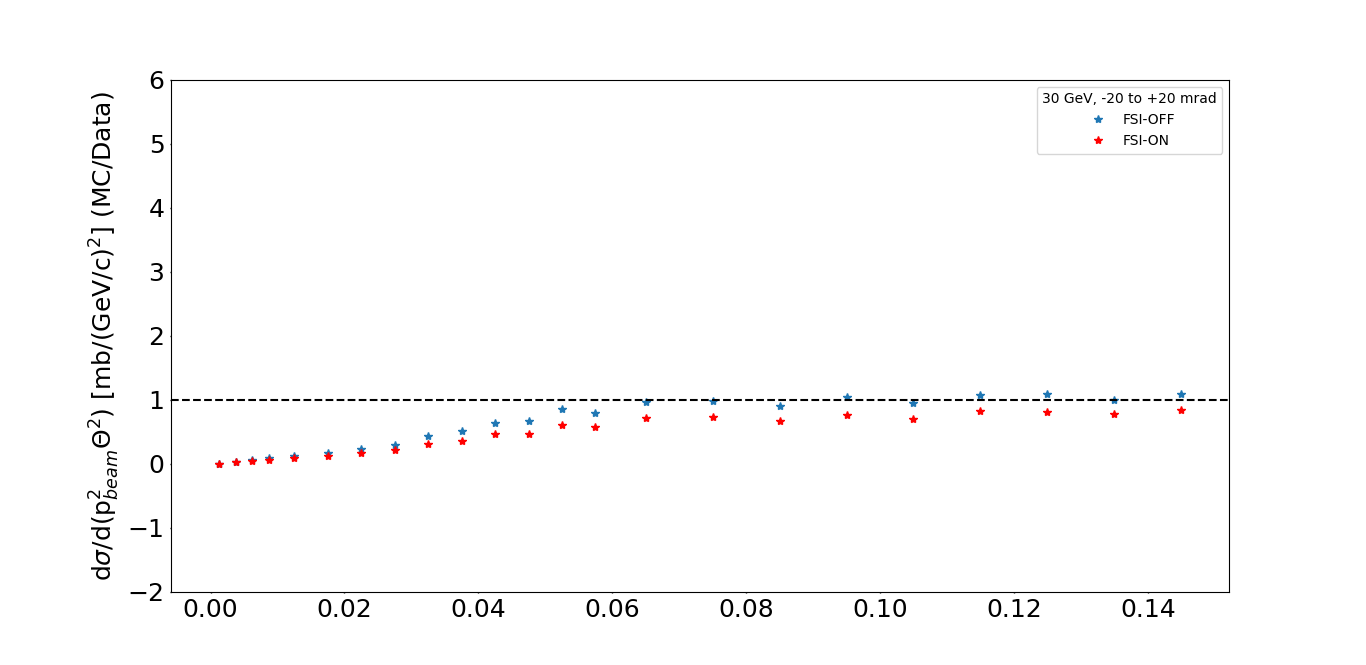} }
	
	\subfloat{\includegraphics[width=0.50\linewidth]{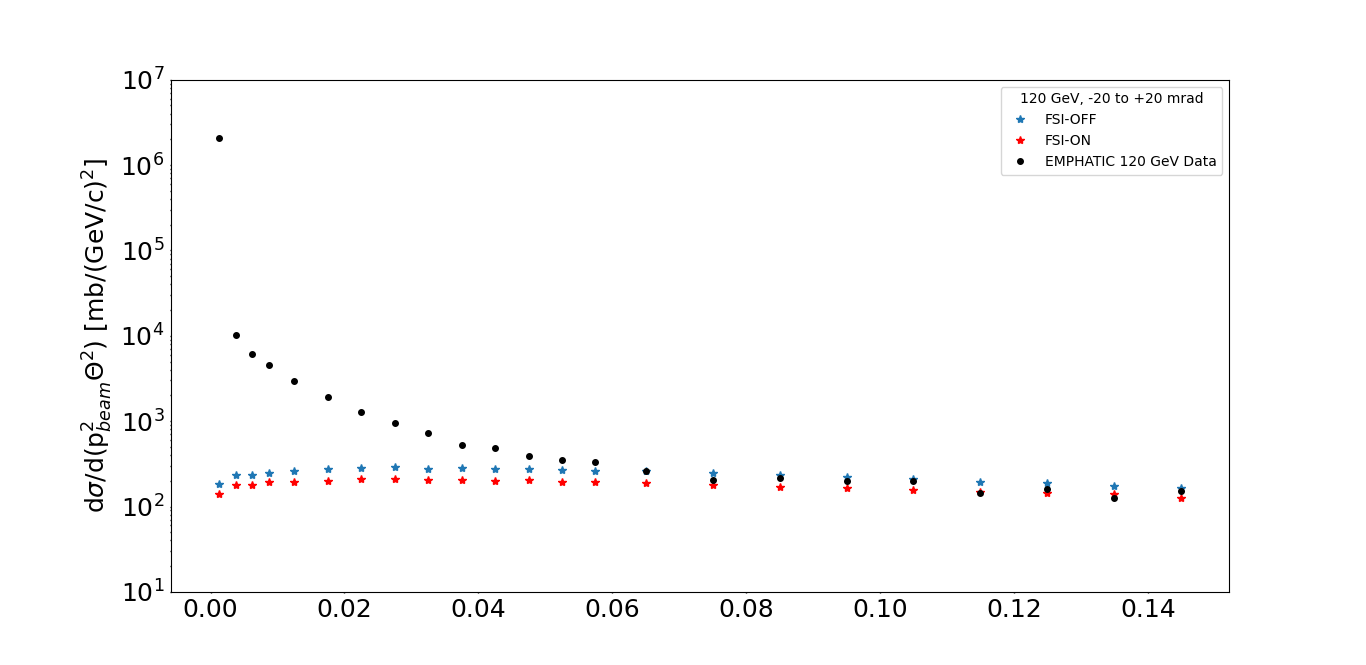} }	
	\subfloat{\includegraphics[width=0.50\linewidth]{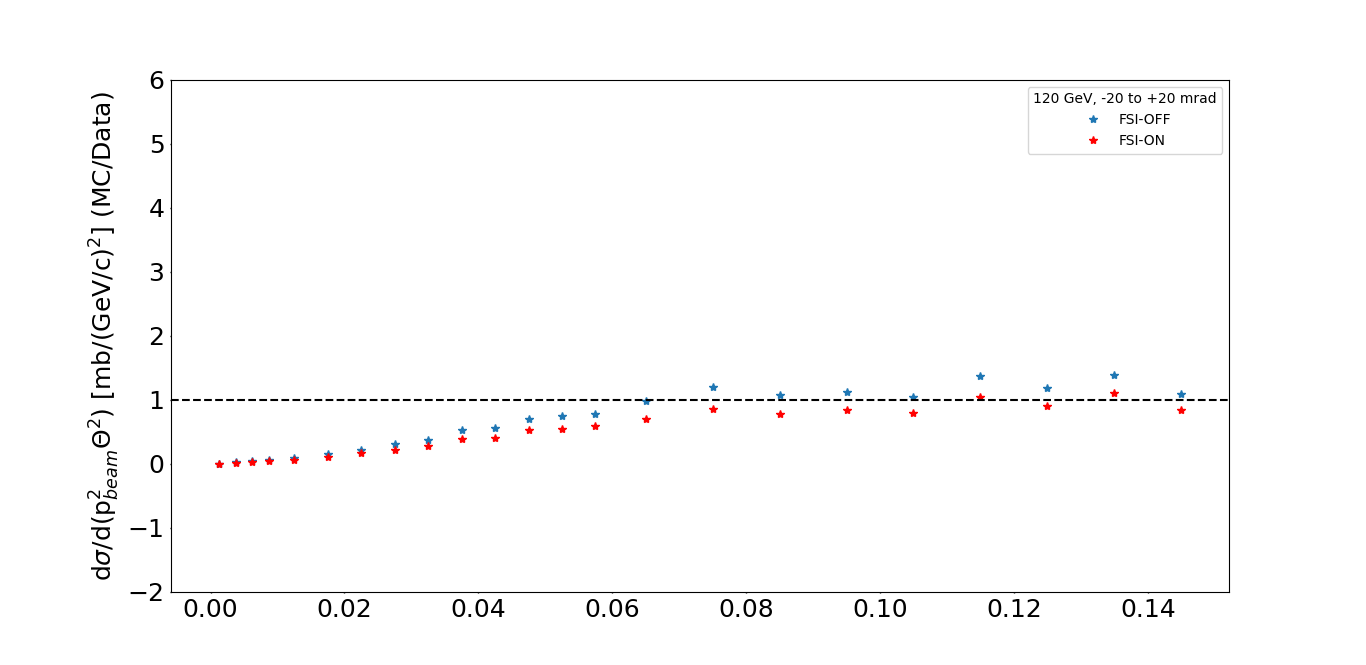} }	
	
	\caption{The plot represents the differential cross-section for a single charge particle in the final state using GiBUU model with FSI-ON(red dotted) and FSI-OFF(blue dotted) compared with EMPHATIC data(black) in the left panel and the right panel shows the ratio of the GiBUU MC with the data}
	\label{fig:7}
\end{figure}

We assess the computations with the incident beam of proton at 20, 30 and, 120 GeV/c  momenta, hitting the Carbon target. The calculated differential cross-section ($\frac{d \sigma}{d(p_{beam}^{2} \theta^{2})}$) for a single charged particle produced within $\pm$20 mrad of the scattering angle is presented in Fig.\ref{fig:7}. Here, our calculations using the GiBUU model show good agreement in comparison with the EMPHATIC data for all beam momenta at higher $p^{2}_{beam}\theta^{2}$($\ge$ 0.045) but could not reproduce the data at lower $p^{2}_{beam}\theta^{2}$($\le$ 0.045). We can observe that GiBUU model is able to predict the data accurately at the higher $p^{2}_{beam}\theta^{2}$ where the quasi-elastic channel (i.e incoherent scattering on the nucleons) dominates while the model could not completely reproduce the region where the coherent elastic scattering dominates.
In our observations, we noticed from Fig.\ref{fig:7} (right panel) that the GiBUU model calculations exhibit a slight preference for the measurement when the effect of the final state interaction is not considered, contrasting with the scenario where the final state interaction is taken into account, especially for the 30 GeV/c case at the higher $p^{2}_{beam}\theta^{2}$.

\section{Conclusions}
\label{sec:6}
The production of charged pions, kaons, and protons at 31 GeV/c is calculated  within the GiBUU model and comparsion with the NA61/SHINE results is shown.
 The agreement obtained for the kaon and pion production is good, mostly for the kaons where we have obtained a good fit across all the scattering angles, and a slight variation for the pions at the backward angles ($\theta$ $>$ 100 mrad). The best fit for the $\pi^{+}$  and $\pi^{-}$ production arise with the treatment of the final state interactions at the angles (20 $< $ $\theta$ $<$ 40 mrad, 300 $< $ $\theta$ $<$ 360 mrad) and (20 $< $ $\theta$ $<$ 40 mrad) respectively.
 Our predictions also agree with the data for the proton production with the best result from the forward angle.  
Furthurmore, the production of the hadrons by proton off the Carbon target using the GiBUU model are compared with the recent EMPHATIC data on single charged particle production on Carbon with proton beams at 20, 30 and 120 GeV/c momenta. Our calculations show very good agreement with the data at the higher $p^{2}_{beam}\theta^{2}$ where the quasi-elastic interaction dominates but contradict the data at lower $p^{2}_{beam}\theta^{2}$ due to the elastic interaction of the proton-nucleus interactions. It is important to note that our predictions for both the EMPHATIC and NA61/SHINE measurements have a correlation that the effect of the final state interaction is insignificant for the proton production as most of the produced particles in the EMPHATIC measruements are protons.

Overall, the GiBUU calculations display better agreement with the measurement when the final state interaction is included except for the proton production, where the model predictions bias the data without the final state interaction. This interestingly could be due to the effect of Color Transparency (CT)  as hinted here\cite{Gallmeister:2022gid,Dutta:2012ii}. However a detailed analysis of this studies is beyond the scope of our current work. 
While the hadron production data are especially useful to the predictions of neutrino flux, they are also beneficial for validating the interactions of neutrinos with nuclei models due to the hadronic re-interaction inside the nuclei. These studies can be useful for improving the GiBUU simulation tool for precision physics.

\section*{Acknowledgements}
RL would like to thank Council of Scientific and Industrial Research, Government of India for the fellowship (file number: 09/1001(0054)/2019-EMR-I). RL and AG would like to thank Department of Science and Technology, Government of India for the financial support through SR/MF/PS-01/2016-IITH/G. We thank Prof. Ulrich Mosel for useful discussions in using GiBUU.
 
 \section*{Data Availability Statement}
All the data that supports the result of this study are included within the article.

\bibliographystyle{apsrev}
	\bibliography{cas-refs}
%

\end{document}